\documentclass[aps,reprint,superscriptaddress,longbibliography]{revtex4-2}
\usepackage{amsmath,amssymb,bm}
\usepackage{graphicx}
\usepackage{xcolor}
\usepackage[normalem]{ulem}
\usepackage{hyperref}

\newcommand{\dd}{\mathrm{d}}
\newcommand{\epsz}{\epsilon_0}
\newcommand{\Acell}{A}
\newcommand{\Rvec}{\bm R}
\newcommand{\Fvec}{\bm F}
\newcommand{\Hmat}{\underline{\bm H}}
\newcommand{\Dmat}{\underline{\bm D}}
\newcommand{\Mmat}{\underline{\bm M}}
\newcommand{\Zt}{\bm{\tilde Z}}
\newcommand{\U}{\phi}
\newcommand{\q}{q}
\newcommand{\SEBEC}{SEBEC}

\newcommand{\atomI}{I}
\newcommand{\Bcond}{\mathcal{B}}
\newcommand{\epsinf}{\underline{\bm \epsilon}_{\infty}}
\newcommand{\Omcell}{V}
\newcommand{\pvec}{\bm p}
\newcommand{\Zstar}{\underline{\bm Z}^{*}}
\newcommand{\Zstara}{\underline{\bm Z}^{*}_{\atomI}}
\newcommand{\Kten}{\underline{\underline{\bm K}}}

\begin{document}

\title{Local Response Theory of Electrified Interfaces from Screened Effective Charges}

\author{Nicolas G. H\"{o}rmann}
\email{hoermann@fhi.mpg.de}
\affiliation{Fritz-Haber-Institut der Max-Planck-Gesellschaft, Faradayweg 4--6, D-14195 Berlin, Germany}

\author{Nicolas Bergmann}
\affiliation{Fritz-Haber-Institut der Max-Planck-Gesellschaft, Faradayweg 4--6, D-14195 Berlin, Germany}

\author{Karsten Reuter}
\affiliation{Fritz-Haber-Institut der Max-Planck-Gesellschaft, Faradayweg 4--6, D-14195 Berlin, Germany}

\begin{abstract}
First-principles, electronic-structure simulations provide direct access to ground-state, global system properties. In the context of computational electrochemistry these are, e.g., bias-dependent energies, free energies, work functions, and capacitances. However, a comparably systematic local description of how these quantities respond to biasing via interfacial charging is still missing.
Here, we introduce the screened electrochemical Born effective charge (\SEBEC), a mixed total-energy derivative that measures the force response of an atomic degree of freedom to interfacial charging under electrochemical boundary conditions. {\SEBEC}s play the role for electrochemistry that Born effective charges play in the modern theory of polarization: they provide an unambiguous decomposition of global response and generate the same formal structures for structural relaxation, capacitance, and electrochemical Stark tuning that are familiar from solid-state response theory. The broader implication is a local response framework that brings electrochemical interfaces closer to the conceptual rigor long established for insulating solids.
\end{abstract}

\maketitle

\section{Introduction}

Electrochemical interfaces are central to technologies for energy storage, electrocatalysis, and electrosynthesis. First-principles electronic-structure theory provides increasingly reliable descriptions of their most relevant properties such as work functions and capacitances as well as reaction energies at applied bias conditions
\cite{ChanNorskov2015Barriers,kastlunger_controlled-potential_2018,Hormann2019GC,Melander2020GCRT, Lindgren2022EGC, Le2020Deciphering,Kronberg2021Reconciling,Li2024Deciphering,Diesen2025Origin,Chen2023Fundamental,Beinlich2023Controlled,HORMANN2025GCreview}. Yet, a rational understanding of the obtained results from local properties remains challenging as the studied interfacial environments often consist of a few tens or hundreds of atoms. Electronic structure and geometry-based descriptors such as d-band centers, local binding and solvation motifs, or H-bond connectivity provide meaningful local features but explain observations typically only at a qualitative level\cite{Shilong2022Descriptors,Hasan2022Understanding,Vallejo2023ABC,Dudzinski2023First,Li2025Kinetic}. Electrostatic models, at variance, can achieve semi-quantitative accuracy by assigning static charges or dipoles to atoms, adsorbates, or reaction coordinates\cite{ChanNorskov2015Barriers,Ge2020On,Le2020Molecular}. This reflects the familiar picture that a species at an electrified interface with charge $Q$ experiences an electrostatic energy contribution $Q\phi$ in a local, average electrostatic potential $\phi$.\cite{Schmickler2010Interfacial,Bazant2013Theory,Boettcher2021Potentially,Hush1958Electrodes,ChanNorskov2015Barriers}. At the atomistic scale, however, both ingredients are nonunique: local charge partitioning is scheme-dependent\cite{MeisterSchwarz1994Ionicity,Anisimov2024Ionicity,Mulliken1955Population,Bader1990AIM,Hirshfeld1977,ReedWeinhold1985NPA}, and there is no universal definition of the electrostatic potential ``experienced'' by a charged particle.\cite{Goldsmith2020Fields,Becker2024_NetzReview,Schmickler2024InnerPotential,Willard2025ElectricFields}

At the same time, an unambiguous local description of electrostatic response has long been established for molecules and insulating bulk systems via dynamical effective charges. In finite molecules, atomic polar tensors (APTs) quantify how displacing each nucleus changes the total molecular dipole moment; equivalently, they identify which nuclei acquire forces when a homogeneous electric field is applied.\cite{PersonNewton1974APT,NewtonPerson1978APT,Cioslowski1989APT} In periodic insulators, Born effective charge (BEC) tensors play the analogous role for the macroscopic polarization: they reconstruct polarization changes by individual atomic displacements and, reciprocally, quantify the field-induced force change for each atom.\cite{KingSmithVanderbilt1993,Resta1994Geometric,GonzeLee1997,Baroni2001Phonons,RestaVanderbilt2007,Spaldin2012} In both cases, the useful descriptor is not an a priori static charge, but a derivative of the global electrostatic response itself. This response-defined character also explains why dynamical charges are central in many electrostatics-aware machine learning frameworks.\cite{schienbein_spectroscopy_2023,falletta_unified_2025,schmiedmayer_derivative_2024,joll_machine_2024,monacelli_electrostatic_2024,gastegger_machine_2021,zhong25a,kim25a}

In the present work, we demonstrate that a systematic local understanding of a range of relevant electrochemical response features can be achieved by considering the atomic force change from biasing an explicitly treated metal-insulator interface with electronic excess charges $q$. While this local response descriptor has already been used in the context of electrochemical barriers\cite{Vijay2022Determining} and for the construction of the bias-sensitive RAZOR machine-learned interatomic potential (MLIP)\cite{bergmann_2025_machine,bergmann_2026_Erasing}, a rigorous discussion of its theoretical foundations and the connections to classical electrostatic response theory are still missing.

As we will show here, the central force-charge derivative naturally leads to an electrochemical analogue of a Born effective charge, which, however, also incorporates metallic and environmental screening. We term this dynamical effective charge the \emph{screened electrochemical Born effective charge} (\SEBEC) and show that {\SEBEC}s generate response structures analogous to those familiar from solid-state polarization theory, as summarized in Table~\ref{tab:ec_vs_ssp}. As a result, our work establishes a local linear-response framework for electrified interfaces that connects directly to the modern theory of polarization for insulating solids, while avoiding the problematic concept of (macroscopic) electrostatic fields at a biased metal-insulator boundary.

\begin{figure*}[t]
\centering
\includegraphics[width=\textwidth]{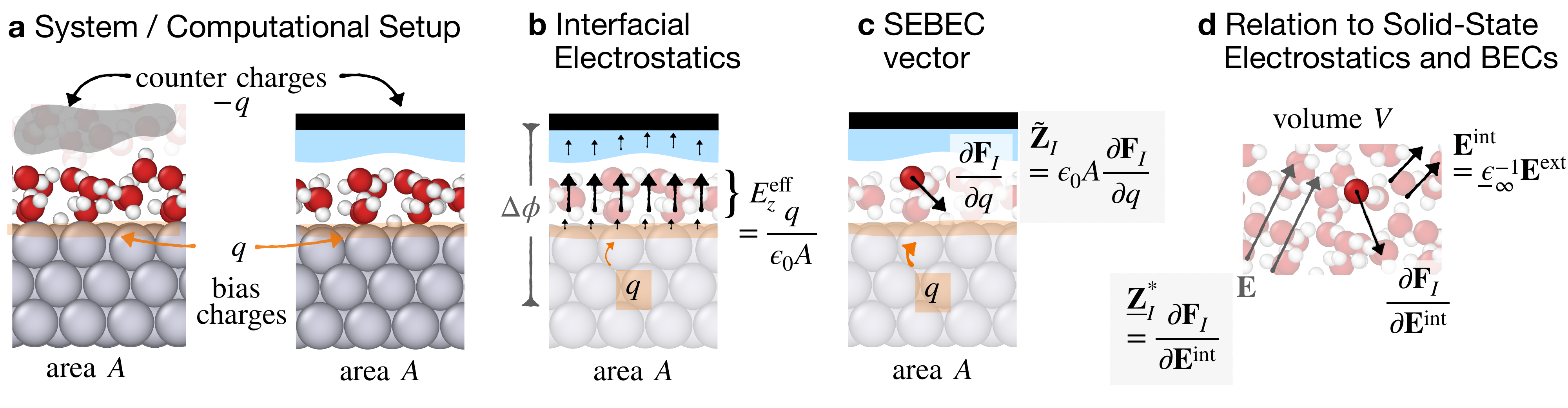}
\caption{(a) Schematic computational setups considered in this work: Interfacial atoms or molecules on a metallic electrode with electrolyte degrees of freedom explicitly integrated out or represented by an implicit environment\cite{Ringe2019Implicit}. (b) Illustration of the effective field concept $E^{\rm eff}_z$ used to define screened electrochemical Born effective charge (\SEBEC) vectors. Actual interfacial electrostatic fields associated with adding excess charges $\q$ are different (see text and Sec. S2 in the Supplemental Material) (c) Definition of {\SEBEC} vectors $\Zt_{\atomI}$ via the force response to interfacial charging. (d) Illustration of the central quantities relevant for description of electrostatic response in solids via Born effective charge tensors $\underline{\bm Z}^{*}_{\atomI}$.}
\label{fig:setup}
\end{figure*}

\section{System Setup and Scope}

Dynamical effective charges in extended systems are response coefficients defined together with the relevant boundary condition. Conventional BECs are transverse, fixed-macroscopic-field coefficients, whereas longitudinal or Callen charges correspond to fixed-displacement-type conditions.\cite{Ghosez1998Dynamical}
The same distinction is central at electrified interfaces, where constant-potential and constant-charge conditions\cite{Hormann2024Converging} (constant Fermi level vs constant electron number) are closest in boundary-condition spirit to fixed-field and fixed-displacement descriptions\cite{MeyerVanderbilt2001SurfacesField,StengelSpaldin2007MetalInsulator,StengelSpaldinVanderbilt2009ElectricDisplacement,WuStengelRabeVanderbilt2008Superlattices}, respectively. Although constant-potential conditions are the most common thermodynamic boundary condition in experiments, they seem less convenient as a starting point for a local response theory: if an atomic displacement changes the potential, a potentiostat must add or remove biasing electrode charges to keep the potential fixed. As a result, the measured response mixes local microscopic rearrangements with electrode-charge readjustments that are strongly system size and setup dependent.

In contrast, canonical, constant-charge simulations fix the electrode charge and thus the interfacial bias fields, which tentatively leads to a more local force response that is less sensitive to boundary conditions (see e.g. discussion in Ref. \onlinecite{Hormann2024Converging}). 

While the present work will therefore only discuss dynamical charges obtained at constant-charge boundary conditions, a corresponding assessment under constant-potential boundary conditions would be an interesting direction for future work.

Figure~\ref{fig:setup}~a summarizes the class of systems considered in this work: metal--insulator interfaces under constant-charge bias conditions, with an explicit interfacial region coupled either to an explicit surrounding medium (e.g. a counter-electrode) or to an effective environment that allows biasing\cite{Ringe2019Implicit}. The essential requirement is not a particular embedding strategy, but the existence of a well-defined system energy $E(\Rvec,\q;\Bcond)$ that depends on nuclear positions $\Rvec\in\mathbb{R}^{3N}$ and bias charge $q$ under a specified set of electrochemical boundary conditions $\Bcond$. These boundary conditions include the cell, composition, electronic-structure setup, electrode/countercharge embedding, and any constraints used to define the biased interface. Unless stated otherwise, all derivatives below are taken at fixed $\Bcond$: derivatives with respect to $\q$ are evaluated at fixed nuclear geometry $\Rvec$, while derivatives with respect to $\Rvec$ are evaluated at fixed charge $\q$. 
It is convenient to use both a global component index $i\in\{1,\dots,3N\}$ for the coordinate vector $\Rvec\in\mathbb{R}^{3N}$ for $N$ atoms and a decomposed index $i=(\atomI,\alpha)$, where $\atomI$ labels the atom and $\alpha\in\{x,y,z\}$ labels the Cartesian direction.

\begin{table*}[t]
\centering
\small
\renewcommand{\arraystretch}{1.05}
\setlength{\tabcolsep}{4pt}
\caption{Comparison of \SEBEC-based electrochemical response relations with standard bulk-solid field-response relations based on Born effective charges and polarization theory. The central mathematical relations are derived in the main text and the Appendix. Both perturbative descriptions are based on linear response and become unreliable once explicit charge-transfer between frontier states becomes important.\cite{Umari2002Ab,Souza2002First,bergmann_2026_Erasing}}
\label{tab:ec_vs_ssp}
\begin{tabular}{p{0.22\textwidth}|p{0.36\textwidth}|p{0.36\textwidth}}
\hline
\hline
\textbf{Phenomenon} & \textbf{Electrified interfaces} & \textbf{Bulk solid insulators} \\
\hline
\hline
Controlling bias variable &
$\q$ = bias charge\newline
$\left(E_z^{\mathrm{eff}} \equiv \frac{\q}{\epsilon_0 A} \right)$ &
$\bm E^{\mathrm{int}}=\epsinf^{-1}\bm E^{\mathrm{ext}}$ = internal field (screened external field)\newline
\cite{GonzeLee1997,RestaVanderbilt2007,Baroni2001Phonons} \\
\hline
Conjugate variable &
$\U = \left(\frac{\partial E}{\partial \q}\right)_{\Rvec,\Bcond}$ = electrode potential &
$\pvec = -\frac{\partial E}{\partial \bm E^{\mathrm{int}}}
= \Omcell \bm P$ = extensive polarization / polarization density times volume\newline
\cite{GonzeLee1997,RestaVanderbilt2007,KingSmithVanderbilt1993,Resta1994Geometric} \\
\hline
Effective charge definition &
$\bm{\tilde Z}_{\atomI}\equiv\left(\frac{\partial \bm F_{\atomI}}{\partial E_z^{\mathrm{eff}}}\right)_{\Rvec,\Bcond}
=
\epsilon_0 A \left(\frac{\partial \bm F_{\atomI}}{\partial \q}\right)_{\Rvec,\Bcond}
=
-\epsilon_0 A \left(\frac{\partial \U}{\partial \bm R_{\atomI}}\right)_{\q,\Bcond}$ \newline
$(\bm{\tilde Z}_{\atomI}\in\mathbb{R}^{3\times 1})$ &
$\Zstara\equiv\frac{\partial \bm F_{\atomI}}{\partial \bm E^{\mathrm{int}}}
=
\frac{\partial \pvec^\top}{\partial \bm R_{\atomI}}
=
\Omcell \frac{\partial \bm P^\top}{\partial \bm R_{\atomI}}$ \newline
$(\Zstara\in\mathbb{R}^{3\times 3})$\newline
\cite{GonzeLee1997,RestaVanderbilt2007,Spaldin2012} \\
\hline
Change of conjugate variable by atomic displacements &
$\dd \phi = -\frac{1}{\epsilon_0 A}\sum_{\atomI} \bm{\tilde Z}_{\atomI}^{\top}\dd\bm R_{\atomI}$ &
$\dd \bm P = \frac{1}{\Omcell}\sum_{\atomI} (\Zstara)^\top\dd\bm R_{\atomI}$\newline
\cite{RestaVanderbilt2007,Spaldin2012} \\
\hline
Bias-/field-induced force changes &
$\dd \bm F_{\atomI}
=
\frac{1}{\epsilon_0 A}\bm{\tilde Z}_{\atomI}\dd \q$ &
$\dd \bm F_{\atomI} = \Zstara\,\dd \bm E^{\mathrm{int}}$\newline
\cite{GonzeLee1997} \\
\hline
Geometric relaxation in linear response &
$\dd \bm R_{\mathrm{relax}}
=
\frac{1}{\epsilon_0 A}\Hmat^{-1}\bm{\tilde Z}\,\dd \q$ &
$\dd \bm R_{\mathrm{relax}}
=
\Hmat^{-1}\Zstar\,\dd \bm E^{\mathrm{int}}$\newline
\cite{Wu2005Systematic,WangVanderbilt2006} \\
\hline
Mode effective charges &
$\tilde Z_{\gamma}=\Zt^\top\bm w_\gamma$ &
$\bm Z^{*}_{\gamma}=(\Zstar)^\top\bm w_\gamma$ \\
\hline
Macroscopic response by relaxation &
$C^{-1}_{\mathrm{relax}}
=
-\frac{1}{(\epsilon_0 A)^2}\bm{\tilde Z}^\top\Hmat^{-1}\bm{\tilde Z}
=
-\frac{1}{(\epsilon_0 A)^2}\sum_{\gamma}\frac{\tilde Z_\gamma^2}{\omega_\gamma^2}$ &
$\bm\chi^{\mathrm{ion}}
=
\frac{1}{\epsilon_0 \Omcell}\sum_{\gamma}
\frac{\bm Z^{*}_{\gamma}\bm Z^{*\,\top}_{\gamma}}{\omega_\gamma^{2}}$\newline
\cite{GonzeLee1997,Baroni2001Phonons,Wu2005Systematic,RestaVanderbilt2007} \\
\hline
Bias-induced vibrational frequency shifts &
$\frac{\partial \omega_\gamma}{\partial q}
\approx
\frac{1}{2\omega_\gamma}
\bm w_\gamma^\top
\left(\frac{\partial \underline{\bm H}}{\partial q}\right)
\bm w_\gamma
$\newline$
\left(\frac{\partial \omega_\gamma}{\partial q}\right)_{\rm direct} \approx
-\frac{1}{2\omega_\gamma\epsz \Acell}\partial_\gamma \tilde Z_\gamma, \quad
\partial_\gamma=\bm w_\gamma^\top\frac{\partial}{\partial \Rvec}$  &
$\frac{\partial \omega_\gamma}{\partial E^{\mathrm{int}}}
\approx
\frac{1}{2\omega_\gamma}
\bm w_\gamma^\top
\left(\frac{\partial \underline{\bm H}}{\partial E^{\mathrm{int}}_k}\right)
\bm w_\gamma $\newline
\cite{WangVanderbilt2006,Souza2002} \\
\hline
\hline
\end{tabular}
\end{table*}

\begin{figure*}[t]
\centering
\includegraphics[width=\textwidth]{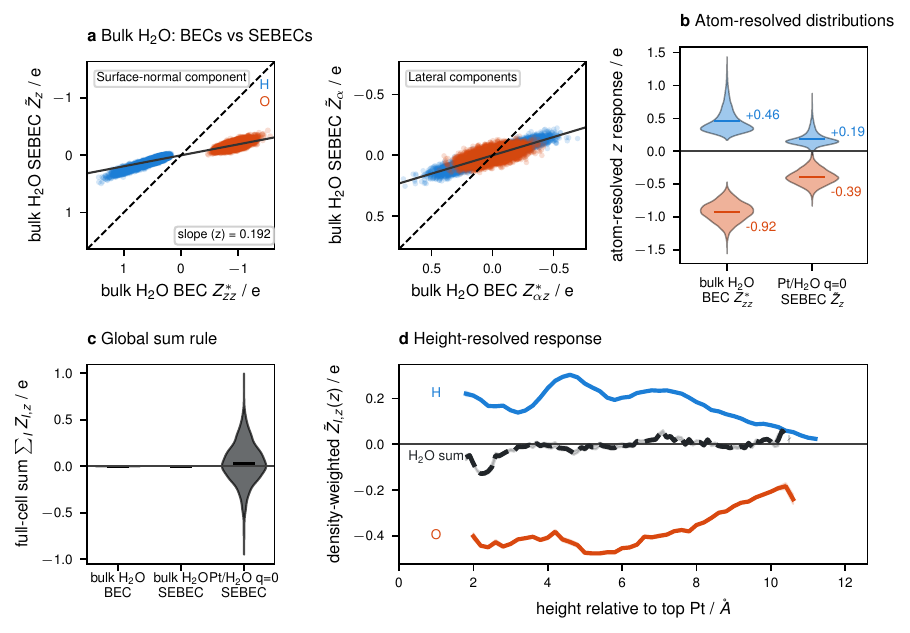}
\caption{BEC--SEBEC comparison for bulk water and a Pt(111)/H$_2$O interface.
(a) Structure-by-structure comparison between reference bulk-water BECs reported in Ref. \onlinecite{schmiedmayer_derivative_2024} and SEBEC components predicted by a Pt/H$_2$O RAZOR model for the same structural dataset. The two subpanels compare the surface-normal responses $Z^{*}_{I,zz}$ and $\tilde Z_{I,z}$ and the lateral $\alpha=x,y$ components $Z^{*}_{I,\alpha z} $ and $\tilde Z_{I,\alpha}$. (b) Atom-resolved distributions of the surface-normal components for the reference bulk-water BECs and own Pt/H$_2$O interfacial-water {\SEBEC} data computed with a DFT-SJM setup. (c) Full-cell sums of the corresponding response objects of panels a and b illustrate that interfacial {\SEBEC}s do not (have to) recover the zero-sum rule (see text). (d) Distance- and atom-resolved mean values $\tilde Z_{I,z}(z)$ for O and H atoms at the Pt/H$_2$O interface; the dashed curve reports the molecular water sum. See SM Sec.~S1 for computational details.
}
\label{fig:response_neutrality}
\end{figure*}

\begin{figure*}[t!]
\centering
\includegraphics[width=0.9\textwidth]{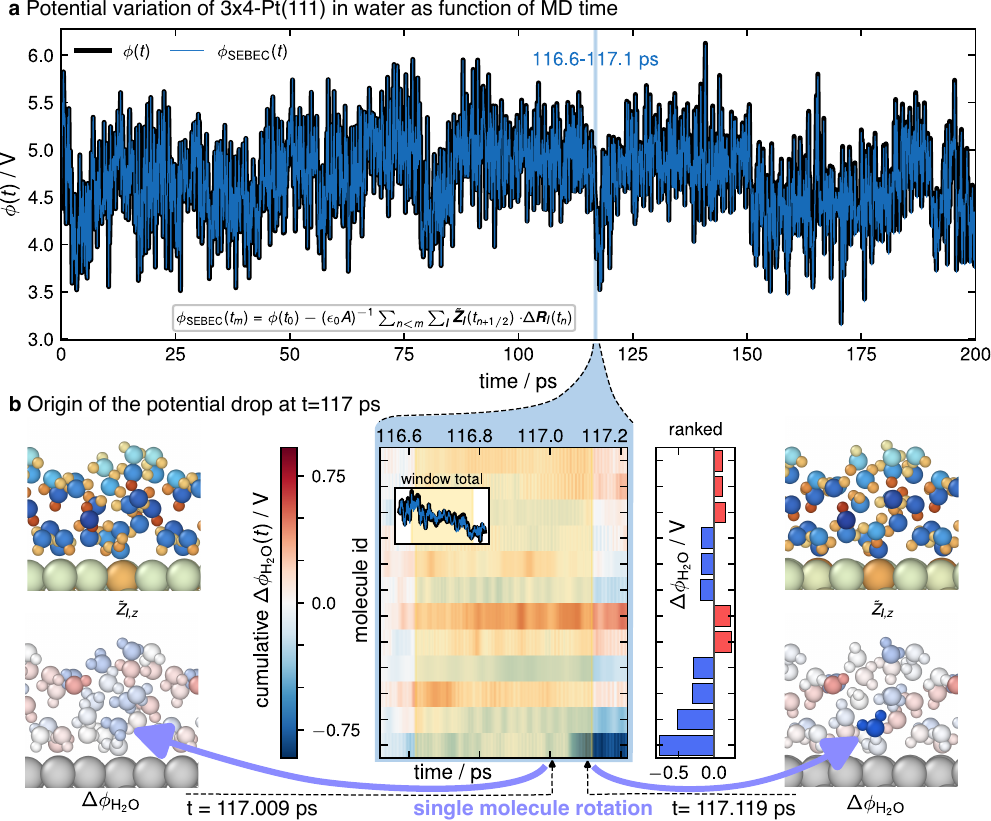}
\caption{Trajectory-level illustration of atom- and molecule-resolved decomposition with {\SEBEC} vectors. (a) Reconstruction of the electrode potential along a $200$ ps, unbiased ($q=0$) Pt(111)-H$_2$O RAZOR-MLIP trajectory using the {\SEBEC}-based, accumulated potential change $\phi_{\mathrm{SEBEC}}(t_m)=\phi(t_0)-(\epsilon_0 A)^{-1}\sum_{n<m}\sum_I \tilde{\bm Z}_{I}(t_{n+\frac12})\cdot\Delta\bm R_I(t_n)$. The highlighted time $116.6$--$117.1$ ps marks a sudden potential drop event, which is analyzed in more detail in panel (b). The central heat map shows cumulative molecular contributions $\Delta\phi_{\mathrm{H_2O}}(t)$ over the highlighted window, ranked by their event-integrated contribution (cf. the adjacent bar plot). The analysis highlights the involvement of multiple interfacial water molecules across the full 0.5 ps time scale. However, the last sudden drop by $\sim -0.7$ V at around $117.05$ ps clearly arises from the reorientation of only a single interfacial water molecule, as underlined by the instantaneous atom-resolved {\SEBEC} values $\tilde{\bm Z}_{I}$ and molecule-resolved potential contributions $\Delta\phi_{\mathrm{H_2O}}(t)$ for two snapshots before and after $t = 117.05$ ps; see SM Sec. S1 for computational details.}
\label{fig:potential_decomposition}
\end{figure*}

\section{Screened Electrochemical Born Effective Charges}

Using this notation, the constant-charge potential-energy surface is $E(\Rvec,\q;\Bcond)$ and the corresponding force vector $\Fvec(\Rvec,\q;\Bcond)\in\mathbb{R}^{3N}$ is
\begin{equation}
\Fvec(\Rvec,\q;\Bcond)=-\left(\frac{\partial E}{\partial \Rvec}\right)_{\q,\Bcond}.
\label{eq:force}
\end{equation}
Componentwise, this means $F_i=-(\partial E/\partial R_i)_{\q,\Bcond}$, or equivalently $F_{\atomI,\alpha}=-(\partial E/\partial R_{\atomI,\alpha})_{\q,\Bcond}$ for $i=(\atomI,\alpha)$ (cf. Fig. ~\ref{fig:setup}~c).

The conjugate intensive variable to bias charge $\q$ is the electrode potential
\begin{equation}
\U(\Rvec,\q;\Bcond)=\left(\frac{\partial E}{\partial \q}\right)_{\Rvec,\Bcond} , 
\label{eq:potential}
\end{equation}
which is a global property directly connected with the Fermi-level position $\tilde{\mu}_{e^-}(\Rvec,\q;\Bcond)$ relative to the appropriate electrostatic reference of the computational setup
\begin{equation}
\U(\Rvec,\q;\Bcond)=-\tilde{\mu}_{e^-}(\Rvec,\q;\Bcond)/e - \U^{\rm ref}(\Bcond) . 
\label{eq:potential_fermilevel}
\end{equation}

The computational setups considered in the numerical analyses of this work are schematically drawn in Fig.~\ref{fig:setup}~b, c. Reported numerical results are either direct outputs of explicit constant-charge DFT calculations coupled to a continuum, implicit solvent and counter-charge model \cite{Andreussi2012Revised,Nattino2018Continuum,Ringe2019Implicit,kastlunger_controlled-potential_2018}, or, from RAZOR MLIP predictions trained on such data\cite{bergmann_2025_machine,bergmann_2026_Erasing} (see Supplemental Material (SM) Sec.~S1 for more computational details). 

In all cases, the electrostatic reference potential $\U^{\rm ref}(\Bcond)$ is evaluated inside the bulk of the implicit solvent which corresponds essentially to potentials $\U(\Rvec,\q;\Bcond)$ on the absolute potential scale\cite{Hormann2019GC}. We note in passing that total energy and Fermi-level-based electrode potential definitions according to Eqs.~(\ref{eq:potential}) and 
(\ref{eq:potential_fermilevel}) intrinsically integrate electrostatic and electronic-structure-related contributions to the potential \cite{Binninger2021Piecewise,Melander2024Constant,Oschinski2024Constant}, highlighting the nuanced differences between modern electronic-structure-based and classical electrostatics-centered discussions of electrified interfaces.

The quantity central to the present work is the mixed derivative of the total energy
\begin{equation}
\left(\frac{\partial^2 E}{\partial \q\,\partial R_i}\right)_{\Bcond}
=
-\left(\frac{\partial F_i}{\partial \q}\right)_{\Rvec,\Bcond}
=
\left(\frac{\partial \U}{\partial R_i}\right)_{\q,\Bcond}.
\label{eq:mixed}
\end{equation}
Equation~(\ref{eq:mixed}) already makes the key point: the same quantity measures how charging changes the force on an atomic degree of freedom and how an atomic displacement changes the electrode potential.

To make contact with the field-response language of solids, it is convenient to map the charging perturbation to an effective interfacial field
\begin{equation}
E_z^{\mathrm{eff}} \equiv \frac{\q}{\epsz \Acell},
\label{eq:effectivefield}
\end{equation}
where $\Acell$ is the lateral interfacial area (cf. Fig. ~\ref{fig:setup}~b). This is a bookkeeping device and intellectual helper rather than a claim that the microscopic field is uniform, unscreened, or structure independent. 
We then define the {\SEBEC} vector
\begin{equation}
\Zt \equiv \left(\frac{\partial \Fvec}{\partial E_z^{\mathrm{eff}}}\right)_{\Rvec,\Bcond}
=
\epsz \Acell\left(\frac{\partial \Fvec}{\partial \q}\right)_{\Rvec,\Bcond}
=
-\epsz \Acell\left(\frac{\partial \U}{\partial \Rvec}\right)_{\q,\Bcond}.
\label{eq:sebec}
\end{equation}
$\Zt$ is therefore a $3N$-component response vector, while the atom-resolved block $\Zt_{\atomI}\in\mathbb{R}^{3}$ measures the (screened) coupling between interface charge $q$ and the three Cartesian force components on atom $\atomI$. In contrast to other works, we do not factor out the elementary charge from this definition, so the effective charges discussed here retain units of charge.
Because $\Zt_{\atomI}$ is a vector response to the scalar charging perturbation, lateral bias-induced forces are included directly in the same ansatz. $\Zt_{\atomI}$ vectors are thus much more expressive than point-charges, and effectively include bias-modified in-plane interactions as well as induced lateral field inhomogeneities (cf. Sec.~S2 and Fig.~S1 in the SM).

In bulk insulators, Born effective charges are conventionally defined with respect to the macroscopic internal field $\bm E^{\mathrm{int}}=\epsinf^{-1}\bm E^{\mathrm{ext}}$ as illustrated in Fig. ~\ref{fig:setup}~d. At electrified interfaces, by contrast, no unique homogeneous internal field exists: the spatial extent of field-creating excess electrons\cite{Li2025Electron} as well as water-structure-dependent variations in high-frequency electronic screening make the local field environment spatially inhomogeneous (cf. Ref. \onlinecite{zhu2025dielectric} and Sec.~S2 in the SM). We therefore avoid introducing an interfacial analogue of $\epsinf$ and use $E_z^{\mathrm{eff}}$ only as a convenient scalar parametrization of the applied charging perturbation.

The analogy of definition (\ref{eq:sebec}) and the Born effective charge (BEC) tensor $\Zstara$ in insulating solids is direct with
\begin{equation}
\Zstara \equiv \frac{\partial \bm F_{\atomI}}{\partial \bm E^{\mathrm{int}}}
=
\frac{\partial \pvec^\top}{\partial \bm R_{\atomI}}
=
\Omcell \frac{\partial \bm P^\top}{\partial \bm R_{\atomI}},
\label{eq:bec}
\end{equation}
with $\bm P$ the polarization density, and $\pvec=\Omcell\bm P$ the corresponding extensive polarization (cf. Figure~\ref{fig:setup}).\cite{GonzeLee1997,RestaVanderbilt2007,Baroni2001Phonons} Here $\Zstara\in\mathbb{R}^{3\times 3}$ is a second-rank tensor because the perturbation itself has three spatial components, whereas the electrochemical quantity $\Zt_{\atomI}\in\mathbb{R}^{3}$ is a vector because the bias perturbation is represented by the single scalar control variable $\q$. The formal structure is therefore closely parallel, while the physics differs in two ways. First, the natural perturbation at an electrified interface is electrode charging rather than a homogeneous bulk field. Second, metallic and environmental electronic screening are folded into the electrochemical response object itself (cf. Figure~\ref{fig:setup}).

Thus, {\SEBEC} vectors are not just projections of BEC tensors onto the surface normal, but screened interfacial response coefficients in their own right. Table~\ref{tab:ec_vs_ssp} makes the formal parallels explicit: the same type of object controls conjugate variables, force changes, structural relaxation, capacitance-like quadratic forms, and vibrational shifts, which we elaborate on below. 

Before, we test the central claim that {\SEBEC}s should be interpreted as BEC-like, screened electrochemical response charges.
For this we compare in Figure~\ref{fig:response_neutrality} bulk-water BECs with {\SEBEC} values for both bulk water and the prototypical Pt(111)/H$_2$O interface. For details on the computational setup, see SM Sec.~S1. Panel~(a) compares published bulk-water BEC elements from Ref. \onlinecite{schmiedmayer_derivative_2024} for oxygen and hydrogen atoms with the respective {\SEBEC} components predicted by a RAZOR ML model trained exclusively on Pt(111)/H$_2$O interface structures evaluated in a DFT-Solvated Jellium Model (SJM)\cite{kastlunger_controlled-potential_2018} setup.\cite{bergmann_2026_Erasing} The near-linear correlation shows that {\SEBEC}s and BECs encode essentially the same response patterns, while slopes below unity are consistent with electronic screening already being integrated into the {\SEBEC} prediction and environment dependence learned by the model. Panel~(b) then compares these bulk-water BEC values with DFT-SJM computed {\SEBEC} values for a Pt(111)/H$_2$O interface, revealing a reduction of the mean effective charge of oxygen and hydrogen atoms by a factor of
$\left\langle Z^*_{zz}\right\rangle_{\rm BEC,bulk} / \langle \tilde{Z}_{z} \rangle_{\rm SEBEC,interface} \sim 2.4$. This magnitude is consistent with literature estimates of the electronic, high frequency dielectric constant in interfacial water ($\epsilon_{\infty}\approx 1.7 ... 2.5$)\cite{zhu2025dielectric} as well as an according analysis of own computations (see SM Sec.~S2 and Figs.~S1--S2\cite{supplemental_material}). In panel~(c), we further study whether the response obeys the zero-sum rule, that is, vanishing center-of-mass forces upon application of a bias. We find that the condition is strictly obeyed for the bulk-water structures, but it is broken for the studied Pt(111)/H$_2$O system. This is unproblematic and in fact expected, because the explicit interfacial subsystem experiences non-vanishing attractive forces from the external counter charge when biased, here represented by the SJM. Finally, panel~(d) shows layer-by-layer oscillations in the mean O and H $\tilde Z_{z}$ profiles, which clarifies that {\SEBEC}s are not invariant static charges but dynamic response coefficients that react to local environment and bonding. The observed oscillations are partly correlated with spatial variations in hydrogen-bond coordination and mean molecular orientations, as shown in SM Sec.~S3 and Fig.~S3.\cite{supplemental_material}. 

Molecular sum values $\tilde Z_{z}(\mathrm{H_2O})$ (dashed line in panel d) are predominantly zero which coincides with torque-only, induced forces upon biasing. Yet, for interfacial water at distances 2 -- 2.5 \AA\ above the Pt surface, ie chemisorbed water\cite{Le2020Molecular}, negative values $\tilde Z_{z}(\mathrm{H_2O})<0$ are observed. These values indicate attractive, center-of-mass forces towards the electrode for positive bias. Indeed, this is consistent with the known stabilization of chemisorbed water for positive bias\cite{Le2020Molecular} and likely derives from partial charge transfer\cite{Le2020Molecular,Khatib2021Nanoscale,Surendralal2021Impact, Li2024Deciphering} and/or electric field gradients in this spatial region (cf. Sec. S2 in the SM). Two examples of atom-resolved variations in $\tilde Z_{z}$ for the Pt(111)/H$_2$O system are provided as well in Fig. ~\ref{fig:potential_decomposition}~b, below.

Taken together, Fig.~\ref{fig:response_neutrality} illustrates that {\SEBEC}s preserve the BEC-like response character while incorporating interfacial screening, metal image-charge effects, boundary conditions, and spatially inhomogeneous fields.

Last, we note that dynamical charges at constant displacement field conditions have been proposed previously for insulating systems\cite{Ghosez1998Dynamical,Martin1981DirectMethod}. These boundary conditions are close in spirit to our constant-charge conditions and these so-called (longitudinal) Callen charges indeed show the same downscaled character as our {\SEBEC}s (cf eq. 9 in Ref. \onlinecite{Ghosez1998Dynamical}). However, due to the unclear nature of boundaries and boundary conditions in our setup with explicitly included metal electrode (and effective electrolyte counter charges) we opt here to use a separate term for the electrochemical response object.

\section{Applications of {\SEBEC}  vectors}

The applications below unpack the most important rows of Table~\ref{tab:ec_vs_ssp}, demonstrating the use of {\SEBEC}s for interpreting potential fluctuations, quasi-static structural response, capacitance, and bias-dependent vibrational behavior within one unified framework. 
\subsection{Unambiguous atom-wise decomposition of interfacial potential drops}

{\SEBEC}  vectors directly link atomic displacements to changes in the interfacial potential drop. The mixed-derivative definition in Eq.~(\ref{eq:mixed}) gives
\begin{align}
\dd\U
&=
\left(\frac{\partial \U}{\partial \Rvec}\right)^\top \dd\Rvec
=
-\frac{1}{\epsz \Acell}\Zt^\top \dd\Rvec \nonumber\\
&=
-\frac{1}{\epsz \Acell}\sum_{\atomI} (\Zt_{\atomI})^\top \dd\bm R_{\atomI}.
\label{eq:dphi}
\end{align}
This converts the global thermodynamic observable central to electrochemistry -- the electrode potential $\U$ -- directly into an atom-resolved signal without assigning static charges to atoms or molecules. Because {\SEBEC}s are vectorial, the same expression also resolves potential changes caused by lateral motion of interfacial species, in contrast to static atomic charges. 

Figure~\ref{fig:potential_decomposition} demonstrates this directly for an unbiased ($q=0$) RAZOR molecular-dynamics trajectory of a Pt(111)/H$_2$O interface\cite{bergmann_2025_machine}.
Panel (a) shows that the {\SEBEC}-reconstructed signal follows the observed potential fluctuations across the full $200$ ps trajectory essentially without drift. Panel (b) then zooms into a representative drop event at $116.6$--$117.1$ ps and visualizes both the instantaneous atom-resolved {\SEBEC} vectors and the corresponding molecule-resolved potential contributions. The dominant signal is traced to the reorientation of a small subset of interfacial water molecules, rather than to a chemisorption event or an indiscriminate collective rearrangement. The highlighted rotation of one single molecule provides the clearest individual contribution towards the end of the jump process. 

In general, potential fluctuations and their relation to interfacial structure and dynamics have been a prominent research area\cite{
Tee2023Constant,Todorova2026First} due to their importance for controlling electrochemical phenomena, e.g. double-layer charging and electrocatalysis. 
While atomic decompositions of global electrostatics are widely established for insulating systems, to the best of our knowledge, the present method provides the first direct demonstration of decomposing interfacial potential dynamics at metal-water interfaces into molecular contributions while mathematically guaranteeing reproduction of the global potential.
Previous approaches to tackle this problem have been rather indirect and approximate as they relied either on the explicit decomposition of the interface structure\cite{Li2024Deciphering} or coupled Wannier-function-based solvent electrostatics with an approximate metallic electrode model\cite{cheng2025} or with an effective, high-frequency screening factor\cite{neugebauer2026}. 

The trajectory-level decomposition in Fig.~\ref{fig:potential_decomposition} and the BEC--SEBEC connection above point toward further dynamical applications, including decompositions of the capacitance and possibly dielectric constants, and spectroscopic observables from interfacial potential (polarization) fluctuations\cite{Tee2023Constant,Staerk2026Static,staerk2026simultaneouslearningstaticdynamic}. 

However, to keep the paper focused on the formal relation between electrochemical and solid-state response summarized in Table~\ref{tab:ec_vs_ssp}, we focus in the following on a linear-response analysis for quasistatic simulations.

\subsection{Structural Relaxation and Screening Contributions to the Capacitance}

Nominally, quasistatic models of electrified interfaces are only appropriate for the description of processes at electrodes interfacing with solid-state electrolytes. Nonetheless, they are equally applied in computational assessments of strongly chemisorbed species and interfacial electrochemical reactions in liquid environments, e.g. water. Despite the apparent lack of realism, the agreement with experiments, in particular trends, is often very good, and according simulations are widely used to rationalize electrochemical phenomena\cite{ChanNorskov2015Barriers,kastlunger_controlled-potential_2018,Lamoureux2019pHEffects,Tiwari2020Fingerprint,Gross2022Reversible,Chen2023Fundamental, Beinlich2023Controlled, Bergmann2023AbInitio}. Within those setups, kinetic properties can be assessed from the potential dependence of barriers, ie the energy difference between local minima and transition states of an underlying potential energy surface. The lowest-order potential dependencies for such stationary states can be rationalized from adsorbate-induced potential changes and capacitances $C$ \cite{Hormann2020Electrosorption,Flores2023Approximating,Beinlich2023Controlled,Oschinski2024On} which are as well affected by adsorbate- and bias-induced structural changes. In this context, the present framework provides a promising route to develop a local rationale for electrochemical function.

To see this more clearly, we reformulate a previously developped linear response logic\cite{Beinlich2023Controlled,Beinlich2023GrandCanonicalPES}, starting from the structural relaxation of a stationary point under a charge perturbation $\dd\q$ (see Appendix~\ref{app:structural_capacitance} for mathematical details) 
\begin{equation}
\dd\Rvec_{\mathrm{relax}}
=
\frac{1}{\epsz \Acell}\Hmat^{-1}\Zt\,\dd\q ,
\label{eq:relax}
\end{equation}
$\Hmat^{-1}$ is the inverse Hessian matrix evaluated for the unbiased situation of interest. These structural changes affect the interfacial potential via Eq.~(\ref{eq:dphi}) and thereby also affect the differential capacitance. The impact can be quantified by writing the inverse capacitance as the sum of a clamped-structure contribution and a relaxation-related contribution
\begin{equation}
C^{-1}
= \frac{\dd \U}{\dd \q} =
\left(\frac{\partial \U}{\partial \q}\right)_{\Rvec}
+
\frac{\partial \U}{\partial\Rvec}^\top
\frac{\partial\Rvec_{\mathrm{relax}}}{\partial q}
\label{eq:ctotal}
\end{equation}
The former term contains the electronic and electrostatic charging response at fixed interface geometry and was previously referred to as $C^{-1}_{\mathrm{el}}$\cite{Beinlich2023GrandCanonicalPES}. The latter relaxation-related term can be analyzed in linear response using Eqs.~(\ref{eq:relax}) and~(\ref{eq:dphi}), giving
\begin{equation}
C^{-1}_{\mathrm{relax}}
\equiv
\frac{\partial \U}{\partial\Rvec}^\top \dd\Rvec_{\mathrm{relax}} =
-\frac{1}{(\epsz \Acell)^2}\Zt^\top \Hmat^{-1}\Zt.
\label{eq:crelax}
\end{equation}
While an according relation was derived previously\cite{Beinlich2023GrandCanonicalPES}, recasting it in terms of {\SEBEC} vectors provides now an atom-resolved picture for the intrinsically global property $C^{-1}_{\mathrm{relax}}$. When combined with atom-decomposed potential variations, one can develop a locally resolved picture for the potential-dependence of stationary points and reaction barriers, which we are exploring in ongoing work and will report on in a separate publication.

In the present study, we rather want to highlight the connections with finite-field lattice dynamics and dielectric response theory.\cite{GonzeLee1997,Wu2005Systematic,WangVanderbilt2006,RestaVanderbilt2007}. For this we reexpress eq.~(\ref{eq:crelax}) in Hessian or dynamical-matrix eigenmodes (see Appendix~\ref{app:structural_capacitance} for details), which yields the compact mode-decomposed form 
\begin{equation}
C^{-1}_{\mathrm{relax}}=
-\frac{1}{(\epsz \Acell)^2}\sum_\gamma \frac{\tilde Z_\gamma^{2}}{\omega_\gamma^{2}} .
\label{eq:crelax_modes}
\end{equation}
$\omega_\gamma$ is the frequency of vibrational mode $\gamma$ and $\tilde Z_\gamma$ the mode effective charge defined by the projection of the {\SEBEC}  vector onto the mass-normalized vibrational mode vector $\bm w_\gamma$: 
\begin{equation}
\tilde Z_\gamma \equiv \Zt^\top \bm w_\gamma. 
\end{equation}

Indeed, the mode-resolved form (\ref{eq:crelax_modes}) for inverse capacitance changes by bias-induced-relaxation directly parallels the ionic contribution to the dielectric susceptibility in solid-state physics\cite{Wu2005Systematic}, as summarized in Table~\ref{tab:ec_vs_ssp}. The equivalence shows that $C^{-1}_{\mathrm{relax}}$ captures classical dielectric-screening physics without invoking electrostatic fields and demonstrates how changes in the charge-potential relation can be rationalized in terms of microscopic contributions without ambiguity. In general we find that soft polar modes -- low-frequency vibrations with large mode effective charge in Eq.~(\ref{eq:crelax_modes}) -- dominate capacitance modifications. 

\subsection{Bias-Dependent Hessians and Electrochemical Stark Tuning}

\begin{figure}[t!]
\centering
\includegraphics[width=0.85\columnwidth]{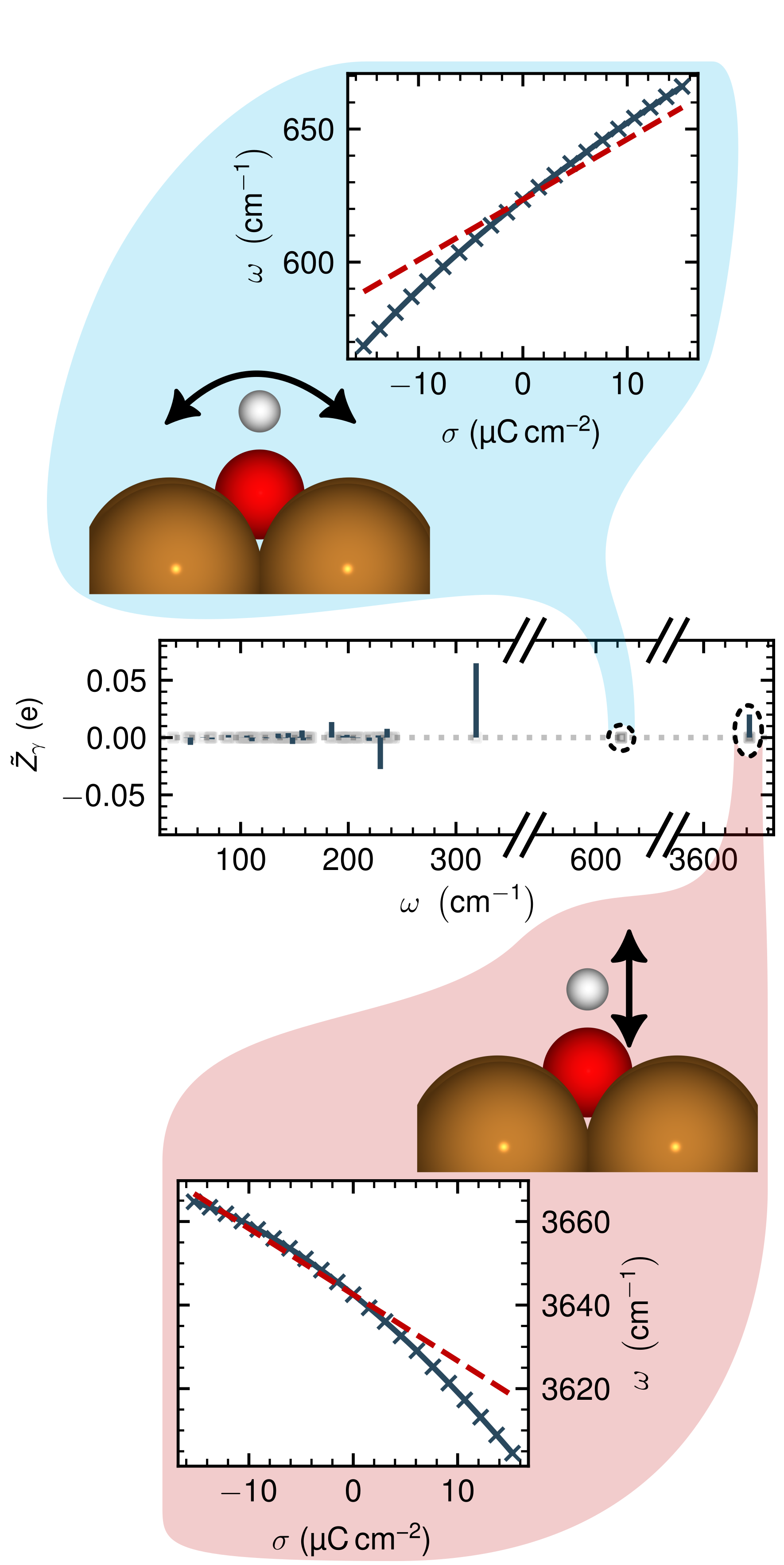}

\caption{Charge dependent vibrational modes for a $p(4\times4)$ OH adsorbed on the fourfold hollow site of the Cu(100) surface, computed with the RAZOR model. In the middle, we plot mode effective charges $\Tilde{Z}_\gamma$ as blue bars against the mode frequencies $\omega$ at zero charge. All mode frequencies $\omega_\gamma$ are in addition indicated by a grey square. In addition, Stark tuning is studied for the two indicated, prototypical modes, OH stretch (bottom, red) and OH rocking (top, blue). The solid lines with markers show vibrational frequencies from explicitly evaluated finite-bias Hessians at applied surface charge densities $\sigma$. The dashed lines refer to predictions via the direct linear-response estimate from the projected {\SEBEC} derivative in Eq.~(\ref{eq:stark}). Offsets and nonlinear deviations are the result of relaxation-induced Hessian changes, anharmonic PES derivatives, and mode mixing, which the linear-response neglects. Notably, the rocking modes have $\tilde Z_\gamma\approx 0$, and therefore nearly vanishing equilibrium field coupling, yet they still exhibit a pronounced Stark response because the latter encodes how the coupling strength varies along the vibrational coordinate.}
\label{fig:stark_demo}
\end{figure}

Charging not only shifts equilibrium positions but also alters the local curvature of the energy landscape. Both effects lead to bias-induced vibrational frequency changes for adsorbates on electrodes -- the electrochemical analogue of vibrational Stark tuning in molecules and field-dependent phonons in solids\cite{Wasileski2001Metal,Wasileski2002Field,Bishop1993vibrational,FriedBoxer2015ACR,Ge2017JPCC,Long2025Impact,DIESEN2025122694,Yang2026Theory}. It can be assessed via perturbation theory (see Appendix~\ref{app:stark_effect}) and knowledge of the charge-induced change of the Hessian 
\begin{equation}
\frac{\dd\Hmat}{\dd q}
=
\left(\frac{\partial \Hmat}{\partial \q}\right)_{\Rvec}
+
\frac{\partial\Hmat}{\partial\Rvec^\top}
\frac{\partial\Rvec_{\mathrm{relax}}}{\partial q}.
\label{eq:si_total_dHdq}
\end{equation}
Using Eq.~(\ref{eq:relax}) for structural response yields then for the relaxation-driven part
\begin{equation}
\frac{\partial\Hmat}{\partial\Rvec^\top}
\frac{\partial\Rvec_{\mathrm{relax}}}{\partial q}
=
\frac{1}{\epsz\Acell}
\Kten\Hmat^{-1}\Zt,
\label{eq:si_dHdq_relax}
\end{equation}
with
\begin{equation}
\Kten \equiv \frac{\partial\Hmat}{\partial\Rvec^\top}
\end{equation}
the third-order force-constant tensor, i.e. the derivative of the Hessian along atomic coordinates. This term therefore requires anharmonic information.  
In contrast, the direct Hessian derivative at the reference geometry can be written as 
\begin{equation}
\left(\frac{\partial \Hmat}{\partial \q}\right)_{\Rvec}
=
-\frac{1}{\epsz \Acell}\frac{\partial \Zt}{\partial \Rvec^\top},
\label{eq:dHdq}
\end{equation}
which connects bias-dependent force constants directly with the geometry dependence of {\SEBEC} values. 

Using perturbation theory and with neglected anharmonic contributions (see Appendix~\ref{app:stark_effect}) one can evaluate the first-order vibrational frequency changes -- the Stark tuning rates -- to be
\begin{equation}
\left(\frac{\partial \omega_\gamma}{\partial q}\right)_{\rm direct}
\approx
-\frac{1}{2\omega_\gamma\epsz \Acell}\partial_\gamma \tilde Z_\gamma, 
\label{eq:stark}
\end{equation}
for vibrational modes $\gamma$ and mode effective charges $\tilde Z_\gamma$ as defined above. 
$\partial_\gamma \tilde Z_\gamma$ represents the derivative of the mode effective charge along the vibrational coordinate,
\begin{equation}
\partial_\gamma=\bm w_\gamma^\top\frac{\partial}{\partial \Rvec} , 
\label{eq:modeprojected_derivative}
\end{equation}
which describes how the coupling strength between bias and vibrational motion changes along the vibrational coordinate.
Hence, the vibrational Stark tuning rates for modes with weak anharmonicity are directly encoded in the mode-projected gradients $\partial_\gamma \tilde Z_\gamma$ of their mode-effective charge. 

The electrochemical Stark literature often considers local field couplings, field-induced changes in bonding or hybridization and electro-inductive through-bond polarization as different mechanisms for vibrational Stark tuning \cite{HEADGORDON199337,Wasileski2001Metal,Wasileski2002Field,Goldsmith2020Fields,Long2025Impact,DIESEN2025122694, Lake2023ElectroInductive,Lake2024El}. 
The derivations here show how all of these mechanisms can be integrated in one coherent description that only includes the geometry dependence of force constants and {\SEBEC} values.

In Figure~\ref{fig:stark_demo} we perform a stringent test for the linear-response description for selected hydroxyl vibrational modes on Cu(100) by comparing the predicted linear frequency shifts for the $\sigma=0$ geometry via Eq.~(\ref{eq:stark}) to explicit dynamical matrix diagonalizations at finite bias. The good overall agreement underlines the importance of the 
leading direct term captured in Eq.~(\ref{eq:stark}), while relaxation and higher-order nonlinearities are only of minor relevance for the selected test cases.

The Stark example closes the sequence of applications that demonstrate that {\SEBEC}s are central components of a local response theory for electrified interfaces within the canonical constant-charge ensemble.

\section{Discussion and Outlook}

The central idea behind the present work is simple: If we are interested in understanding how an applied bias affects the microscopic behavior at electrified interfaces in a constant-charge ensemble, we should declare the differential force change on an atom as the relevant descriptor. As the so-defined {\SEBEC} vector $\Zt_{\atomI}$ equally encodes the linear change of the potential with respect to atomic displacements, it naturally becomes the central object of linear-response analyses of relevant electrochemical observables.
As demonstrated in the previous sections, {\SEBEC}s are the natural electrochemical analogue of traditional Born effective charges and lead to mathematical structures closely paralleling those established for polarization-related phenomena in the solid-state literature, as summarized in Table~\ref{tab:ec_vs_ssp}. They thus bring electrochemical response formulations closer to the rigor and compactness long achieved in solid-state physics and naturally incorporate effects from bond renormalization and electro-induction. At the same time they avoid detours to classical electrostatics and associated conceptual complications arising from inhomogeneous electronic screening and definitions of polarization and field strengths in systems with partially metallic character.

{\SEBEC}s provide genuine local information from which the global response can be reconstructed identically, and, vice versa, allow to decompose global observations systematically into atom-, molecule-, layer-, or mode-resolved contributions. The demonstrated atom-level decomposition of electrode potential variations along an MD trajectory and of the inverse capacitance into mode-resolved contributions are two examples of this. As already mentioned, we anticipate that analogous decompositions of capacitive response also for dynamically evolving interfaces from potential fluctuations are possible \cite{TeeSearles2023,Staerk2026Static} as well as the assessment of spectroscopic properties, e.g. IR intensities (cf. the observed linear correlation between traditional BECs and {\SEBEC}s Fig.~\ref{fig:response_neutrality}, and a brief discussion in SM Sec.~S4.\cite{supplemental_material}).

Finally, we want to clarify that electrochemistry is usually formulated in terms of electrode potentials, a perspective that has naturally driven much recent theoretical development toward constant-potential methods and descriptions.\cite{Hormann2019GC,Melander2020GCRT,Lindgren2022EGC} This creates no formal obstacle, since the switching routines between constant-charge and constant-potential descriptions via a Legendre transform are largely established. Within the present linearized setting, constant-potential derivatives typically follow from the constant-charge response together with the corresponding charge-to-potential conversion, i.e. through straightforward capacitance renormalizations.\cite{Ringe2019CO}
At the same time, the present work equally serves as a blueprint to directly derive dynamical charges and response coefficients in a constant-potential ensemble\cite{Beinlich2023GrandCanonicalPES}, which would be the natural next step to complement the present constant-charge framework.

\begin{acknowledgments}
We thank Christian Carbogno for helpful discussions on density-functional perturbation theory, the modern theory of polarization, and Born effective charges. Furthermore, we acknowledge the German Research Foundation for funding via the DFG Cluster of Excellence e-conversion EXC 2089/1 and the Max Planck Computing and Data Facility (MPCDF) for providing computational resources.
\end{acknowledgments}

\section*{Data Availability}
All relevant data and code needed to reproduce the content of the figures will be made available upon acceptance.

\section*{AI Use Statement}
AI-assisted tools were used to support analysis code development, documentation, and language editing.

\appendix
\section{Mathematical Details of the SEBEC Response Framework}
\subsection{Structural Relaxation and Capacitance}
\label{app:math_details}
\label{app:structural_capacitance}

The starting point for the linear-response discussion is a stationary reference geometry of the constant-charge energy $E(\Rvec,\q;\Bcond)$ with Hessian
\begin{equation}
\Hmat =
\frac{\partial^2 E}{\partial \Rvec\,\partial \Rvec^\top}.
\label{eq:app_hessian}
\end{equation}
At the stationary point, a small charging perturbation $\dd q$ generates the bias-induced force
\begin{equation}
\dd\Fvec_{\mathrm{bias}}
=
\left(\frac{\partial\Fvec}{\partial q}\right)_{\Rvec,\Bcond}\dd q
=
\frac{1}{\epsz\Acell}\Zt\,\dd q .
\label{eq:app_bias_force}
\end{equation}
Force balance at the displaced stationary point gives
\begin{equation}
\Hmat\,\dd\Rvec_{\mathrm{relax}}
=
\dd\Fvec_{\mathrm{bias}},
\label{eq:app_force_balance}
\end{equation}
and therefore the linear structural response
\begin{equation}
\dd\Rvec_{\mathrm{relax}}
=
\frac{1}{\epsz \Acell}\Hmat^{-1}\Zt\,\dd q .
\label{eq:app_relax}
\end{equation}
Combining this displacement with the local potential-decomposition formula,
\begin{equation}
\dd\U
=
-\frac{1}{\epsz\Acell}\Zt^\top\dd\Rvec ,
\label{eq:app_dphi}
\end{equation}
gives the relaxation-induced potential change
\begin{equation}
\dd\U_{\mathrm{relax}}
=
-\frac{1}{(\epsz\Acell)^2}
\Zt^\top\Hmat^{-1}\Zt\,\dd q .
\label{eq:app_dphirelax}
\end{equation}
The corresponding relaxation contribution to the inverse extensive capacitance of the simulated interface cell is
\begin{equation}
C^{-1}_{\mathrm{relax}}
\equiv
\frac{\dd\U_{\mathrm{relax}}}{\dd q}
=
-\frac{1}{(\epsz\Acell)^2}
\Zt^\top\Hmat^{-1}\Zt .
\label{eq:app_crelax}
\end{equation}

To expose the role of collective modes, diagonalize the Hessian as
\begin{equation}
\Hmat\bm u_\lambda=\kappa_\lambda\bm u_\lambda,
\qquad
\Hmat^{-1}=
\sum_\lambda
\frac{1}{\kappa_\lambda}
\bm u_\lambda\bm u_\lambda^\top .
\label{eq:app_hessian_spectral}
\end{equation}
Substitution into Eq.~(\ref{eq:app_crelax}) gives
\begin{equation}
C^{-1}_{\mathrm{relax}}
=
-\frac{1}{(\epsz\Acell)^2}
\sum_\lambda
\frac{\tilde Z_\lambda^2}{\kappa_\lambda},
\qquad
\tilde Z_\lambda\equiv\Zt^\top\bm u_\lambda .
\label{eq:app_hessian_mode}
\end{equation}
Thus a mode contributes strongly when it is both mechanically soft and electrochemically active.  For comparison with standard vibrational notation, introduce the mass matrix $\Mmat$ and dynamical matrix
\begin{equation}
\Dmat=\Mmat^{-1/2}\Hmat\Mmat^{-1/2}.
\label{eq:app_dynamical_matrix}
\end{equation}
The dynamical-matrix eigenvectors $\bm v_\gamma$ satisfy
\begin{equation}
\Dmat\bm v_\gamma=\omega_\gamma^2\bm v_\gamma,
\qquad
\bm v_\gamma^\top\bm v_{\gamma'}=\delta_{\gamma\gamma'} .
\label{eq:app_dmat_eigen}
\end{equation}
The corresponding physical, mass-weighted displacement patterns are
\begin{equation}
\bm w_\gamma=\Mmat^{-1/2}\bm v_\gamma , 
\label{eq:app_w_modes}
\end{equation}
and solve the generalized eigenvalue problem
\begin{equation}
\qquad
\Hmat\bm w_\gamma=\omega_\gamma^2\Mmat\bm w_\gamma .
\end{equation}

Using $\Hmat^{-1}=\Mmat^{-1/2}\Dmat^{-1}\Mmat^{-1/2}$ and
\begin{equation}
\Dmat^{-1}=
\sum_\gamma
\frac{\bm v_\gamma\bm v_\gamma^\top}{\omega_\gamma^2},
\label{eq:app_dmat_inverse}
\end{equation}
one obtains
\begin{equation}
\Zt^\top\Hmat^{-1}\Zt
=
\sum_\gamma
\frac{(\Zt^\top\bm w_\gamma)^2}{\omega_\gamma^2}.
\label{eq:app_quadratic_modes}
\end{equation}
Hence, by defining the vibrational mode effective charge
\begin{equation}
\tilde Z_\gamma\equiv\Zt^\top\bm w_\gamma ,
\label{eq:app_mode_charge}
\end{equation}
the relaxation capacitance becomes
\begin{equation}
C^{-1}_{\mathrm{relax}}
=
-\frac{1}{(\epsz\Acell)^2}
\sum_\gamma
\frac{\tilde Z_\gamma^2}{\omega_\gamma^2}.
\label{eq:app_vibcap}
\end{equation}
In this representation, $C^{-1}_{\mathrm{relax}}$ decomposes into individual mode contributions 
\begin{equation}
C^{-1}_{\gamma}  = 
-\frac{1}{(\epsz\Acell)^2}
\frac{\tilde Z_\gamma^2}{\omega_\gamma^2}.
\end{equation}
with
\begin{equation}
C^{-1}_{\mathrm{relax}}
=
\sum_\gamma
C^{-1}_{\gamma} \ .
\end{equation}

The prefactor $\frac{1}{(\epsz\Acell)^2}$ reflects that the electrochemical response is written for the conjugate pair $(q,\phi)$ of the finite interface cell and the fact that the present work reports inverse capacitances $C^{-1}$ that are inversely extensive with respect to the interface area $\Acell$. In contrast, the bulk ionic susceptibility in insulating solids\cite{GonzeLee1997,RestaVanderbilt2007}, which reads
\begin{equation}
\bm\chi^{\mathrm{ion}}
=
\frac{1}{\epsz\Omcell}
\sum_\gamma
\frac{\bm Z^{*}_{\gamma}\bm Z^{*\,\top}_{\gamma}}{\omega_\gamma^2},
\label{eq:app_chiion}
\end{equation}
is volume-normalized and thus exhibits an accordingly different prefactor.

\subsection{Electrochemical Stark effect}
\label{app:stark_effect}
The electrochemical Stark response is connected with the first-order Hessian change 
\begin{equation}
\Delta\Hmat \approx \frac{\dd\Hmat}{\dd q} \dd q
\label{eq:total_deltaH}
\end{equation}
which maps to a dynamical-matrix change 
\begin{equation}
\Delta\Dmat=
\Mmat^{-1/2}\Delta\Hmat\Mmat^{-1/2}.
\label{eq:app_delta_dmat}
\end{equation}
Non-degenerate, first-order perturbation theory gives the perturbed squared frequency as
\begin{equation}
(\omega_\gamma+\Delta\omega_\gamma)^2
=
\bm v_\gamma^\top
(\Dmat+\Delta\Dmat)
\bm v_\gamma
=
\bm w_\gamma^\top
(\Hmat+\Delta\Hmat)
\bm w_\gamma.
\label{eq:app_freq2_shift}
\end{equation}
Subtracting the unperturbed relation $\omega_\gamma^2=\bm w_\gamma^\top\Hmat\bm w_\gamma$ and linearizing the left-hand side gives
\begin{equation}
2\omega_\gamma\Delta\omega_\gamma
\approx
\bm w_\gamma^\top
\Delta\Hmat
\bm w_\gamma .
\label{eq:app_freq_shift}
\end{equation}
Combined with Eq.~(\ref{eq:total_deltaH}), this yields the central linear frequency derivative
\begin{equation}
\frac{\partial\omega_\gamma}{\partial q}
=
\frac{1}{2\omega_\gamma}
\bm w_\gamma^\top
\frac{\dd\Hmat}{\dd q}
\bm w_\gamma .
\label{eq:app_freq_derivative}
\end{equation}
Equations~(\ref{eq:total_deltaH})--(\ref{eq:app_freq_derivative}) are the general linear-response recipe: compute the Hessian derivative, project the resulting dynamical-matrix change onto the unperturbed mode, and divide by $2\omega_\gamma$. Possible mode mixing effects are therefore not included in these first-order perturbation-theory expressions.

As elaborated in the main text, $\frac{\dd\Hmat}{\dd q}$ can be separated into a direct fixed-geometry term and a relaxation-induced term (eqs. (\ref{eq:si_total_dHdq}) -- (\ref{eq:dHdq}) in the main text). 
\begin{equation}
\frac{\dd\Hmat}{\dd q}
=
\left(\frac{\partial\Hmat}{\partial q}\right)_{\Rvec}
+
\frac{\partial\Hmat}{\partial\Rvec^\top}
\frac{\partial\Rvec_{\mathrm{relax}}}{\partial q}.
\label{eq:app_total_dHdq}
\end{equation}
Keeping only the direct fixed-geometry contribution allows to derive thus the leading-order Stark tuning rate to
\begin{equation}
\left(\frac{\partial\omega_\gamma}{\partial q}\right)_{\mathrm{direct}}
=
-\frac{1}{2\omega_\gamma\epsz\Acell}
\bm w_\gamma^\top
\frac{\partial\Zt}{\partial\Rvec^\top}
\bm w_\gamma .
\label{eq:app_domegadq_direct}
\end{equation}
With the fixed-mode directional derivative
\begin{equation}
\partial_\gamma=
\bm w_\gamma^\top
\frac{\partial}{\partial\Rvec},
\label{eq:app_directional_derivative}
\end{equation}
this is equivalently
\begin{equation}
\left(\frac{\partial\omega_\gamma}{\partial q}\right)_{\mathrm{direct}}
\approx
-\frac{1}{2\omega_\gamma\epsz\Acell}
\partial_\gamma\tilde Z_\gamma .
\label{eq:app_stark_direct}
\end{equation}
The relaxation correction follows analogously as
\begin{equation}
\left(\frac{\partial\omega_\gamma}{\partial q}\right)_{\mathrm{relax}}
=
\frac{1}{2\omega_\gamma\epsz\Acell}
\bm w_\gamma^\top
\left(\Kten\Hmat^{-1}\Zt\right)
\bm w_\gamma ,
\label{eq:app_domegadq_relax}
\end{equation}
and accounts for curvature changes sampled along the bias-induced relaxation path. An instructive one-dimensional oscillator model illustrating the same separation into direct and relaxation/anharmonic contributions is given in Sec.~S5 of the SM.

\bibliographystyle{apsrev4-2}
\bibliography{bib1}

\end{document}